\documentclass[twocolumn,pra,aps,letterpaper,amsmath,superscriptaddress]{revtex4-1}

\usepackage{graphicx}
\usepackage{xspace} 

\newcommand{\tc}{\ensuremath{T_\mathrm{c}}\xspace}
\newcommand{\jc}{\ensuremath{J_\mathrm{c}}\xspace}
\newcommand{\jd}{\ensuremath{J_\mathrm{d}}\xspace}

\newcommand{\hctwo}{\ensuremath{H_\mathrm{c2}}\xspace}

\newcommand{\dg}{\ensuremath{^\circ}\xspace}

\newcommand{\alo}{Al$_2$O$_3$}

\begin{document}

\title{Depairing critical current achieved in superconducting thin films with through-thickness arrays of artificial pinning centers}
\author{Rafael B. Dinner}
\altaffiliation{Present address: Aret\'e Associates, 9301 Corbin Ave, Northridge, CA 91324, USA}
\email[Email: ]{rdinner@stanfordalumni.org}
\affiliation{Department of Materials Science and Metallurgy, University of Cambridge, Pembroke Street, Cambridge CB2 3QZ, UK}
\author{Adam P. Robinson}
\affiliation{Nanoscience Centre, University of Cambridge, 11 J J Thomson Avenue, Cambridge CB3 0FF, UK}
\author{Stuart C. Wimbush}
\affiliation{Department of Materials Science and Metallurgy, University of Cambridge, Pembroke Street, Cambridge CB2 3QZ, UK}
\author{Judith L. MacManus-Driscoll}
\affiliation{Department of Materials Science and Metallurgy, University of Cambridge, Pembroke Street, Cambridge CB2 3QZ, UK}
\author{Mark G. Blamire}
\affiliation{Department of Materials Science and Metallurgy, University of Cambridge, Pembroke Street, Cambridge CB2 3QZ, UK}

\date{\today}

\begin{abstract}
Large area arrays of through-thickness nanoscale pores have been milled into superconducting Nb thin films via a process utilizing anodized aluminum oxide thin film templates. These pores act as artificial flux pinning centers, increasing the superconducting critical current, \jc{}, of the Nb films. By optimizing the process conditions including anodization time, pore size and milling time, \jc{} values approaching and in some cases matching the Ginzburg-Landau depairing current of 30 MA/cm$^2$ at 5~K have been achieved -- a \jc{} enhancement over as-deposited films of more than 50 times. In the field dependence of \jc{}, a matching field corresponding to the areal pore density has also been clearly observed. The effect of back-filling the pores with magnetic material has then been investigated. While back-filling with Co has been successfully achieved, the effect of the magnetic material on \jc{} has been found to be largely detrimental compared to voids, although a distinct influence of the magnetic material in producing a hysteretic \jc{} versus applied field behavior has been observed. This behavior has been tested for compatibility with currently proposed models of magnetic pinning and found to be most closely explained by a model describing the magnetic attraction between the flux vortices and the magnetic inclusions.
\end{abstract}


\maketitle

\section{\label{intro}Introduction}

\begin{figure}
  \includegraphics[width=3.4in]{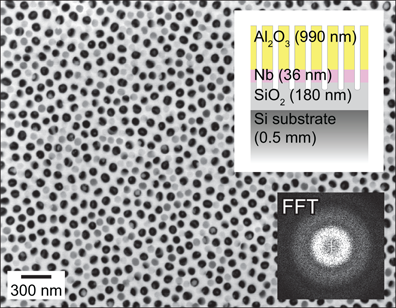}
  \caption{\label{fig_film}(Color online) Anodized aluminum oxide (AAO) forms nanoscale pores over large areas, as shown in this scanning electron micrograph of the surface of an AAO film on Nb (schematic, upper inset). The image shows a disordered, natural arrangement of pores of well defined diameter (75 $\pm$ 10~nm). The rings in the two-dimensional Fourier transform of the image (lower inset) indicate a lack of crystalline order, but a defined pore spacing ($\sim$140~nm).}
\end{figure}

In recent years, the state-of-the-art critical current density, \jc{}, of superconducting films has improved significantly as researchers have developed nanocomposites incorporating artificial pinning centers~\cite{HTS_review}. Non-superconducting inclusions, such as insulating particles within a YBa$_2$Cu$_3$O$_{7-\delta}$ matrix~\cite{bzo,tantalates}, serve to pin magnetic flux. Typically, such inclusions are incorporated randomly into the superconducting matrix, without the fine control of size and spacing that is necessary for optimization of \jc{}. The first part of this work describes a versatile method for fabricating artificial pinning centers with a highly controlled geometry: nanoscale pores, shown in Fig.~\ref{fig_film}, are transferred from an anodized aluminum oxide (AAO) template into a superconducting film. Anodization of alumina is a well-studied and reliable self-assembly process~\cite{aao1, aao_theory, Adam_AAO, Mingzhe_AAO} that can create billions of nanoscale pores in parallel over a large area, with control over both the spacing and size of the pores. Such a scalable process could be integrated into the commercial production of long lengths of superconducting wire~\cite{HTS_review}, unlike slow, serial patterning processes such as electron beam lithography and focused ion beam patterning.

The size and spacing of AAO pores is well matched to the vortex lattice that forms in practical type-II superconductors. The spacing can be varied from at least 300~nm to below 50~nm, corresponding to the spacing of the vortex lattice at fields from 30~mT to 1~T. There have been several efforts to exploit this natural match between AAO and the vortex lattice, including depositing superconducting niobium (Nb) on top of AAO~\cite{Welp}, depositing NbN on wires templated by AAO~\cite{Hallet}, and using AAO as a mask to ion mill holes in Nb~\cite{Keay}. Here, we further develop this last process, creating very strong pinning centers by ensuring through-thickness holes in the Nb layer, and improving the process reliability through secondary ion mass spectrometry during milling.

Having created strong and well-characterized pinning centers in the form of holes in Nb, we then address the basic scientific question of whether \textit{magnetic} pinning centres can improve on their non-magnetic equivalents, a source of significant debate given the importance of progressing beyond the capabilities presently offered by core pinning approaches~\cite{BES}. The AAO system is ideal for backfilling the holes with magnetic material, leading to through-thickness magnetic inclusions that can be directly compared to their non-magnetic counterpart. While various mechanisms have been proposed for magnetic pinning~\cite{SF_hybrids,milosevic,RDS_math,cardoso,Palau}, the most technologically promising~\cite{Blamire_magnetic_pinning_theory} requires such through-thickness inclusions, rather than magnets placed above or below a superconducting film, as in most of the experiments conducted thus far~\cite{velez_review,snezhko,lange,morgan}. In previous experiments with through-thickness magnetic inclusions~\cite{NbDy,Stuart_magnetic_pinning,nanorods}, such inclusions inevitably affect (and usually increase) the core pinning of vortices in the film, making it impossible to isolate the contribution of magnetic interactions to \jc{}. In the system presented here, we overcome this difficulty by changing the magnetization of through-thickness inclusions \textit{in situ} and observing the resultant change in \jc{}. This relationship between pinning and magnetization is manifested as a novel, hysteretic behavior of \jc{} under an applied field.

\section{Nanofabrication of Artificial Pinning Sites}

\begin{figure*}
  \includegraphics[width=6in]{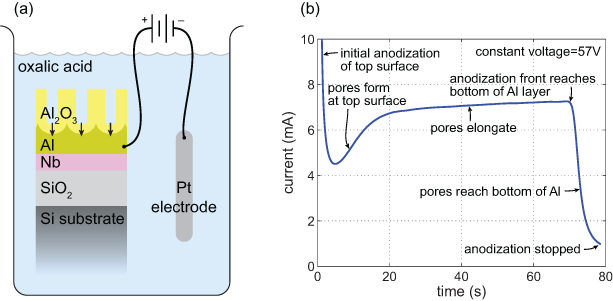}
  \caption{\label{fig_anodization}(Color online) (a) Schematic of the anodization process. The film's cross section (not to scale) is depicted immersed in an oxalic acid bath at a fixed potential, forcing oxidation of the Al and pore formation. The current measured during this process, shown in (b), reveals the various phases of pore formation, which self-terminates at the Nb layer.}
\end{figure*}

The cross-sectional structure of a typical film is illustrated schematically in Fig.~\ref{fig_anodization}(a). A 36~nm thick superconducting niobium (Nb) layer and a 670~nm thick aluminum (Al) layer are deposited in a single sputtering run by moving the oxidized silicon substrate past two DC magnetrons. The chamber is baked out before deposition and liquid nitrogen cooled during deposition, resulting in a background pressure of less than $10^{-8}$~mbar. The aluminum layer is then anodized into \alo{} in 0.3~M oxalic acid at room temperature following a procedure for forming porous AAO on silicon~\cite{Adam_AAO}. During anodization, the substrate and film edges are protected from the acid bath with a temporary adhesive (glycol phthalate).

\begin{figure}
  \includegraphics[width=3in]{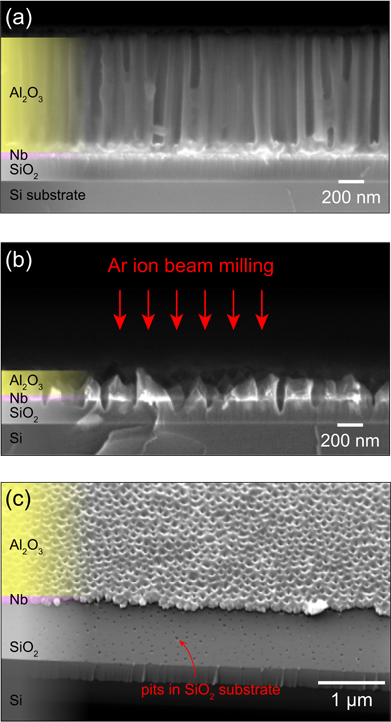}
  \caption{\label{fig_mill}(Color online) SEM images illustrate the pore transfer process. (a) An image of the cleaved edge of the film before milling. (b) After milling, the \alo{} layer is thinned and the pores extend through the Nb layer. (c) Peeling away the Nb and \alo{} layers reveals pits in the SiO$_2$ substrate layer, proving that the pores extend through the Nb.}
\end{figure}

As shown in Fig.~\ref{fig_anodization}(b), the anodization process is self-terminating. A constant voltage is applied, and the current indicates the rate of anodization. The anodization rate is initially high but drops rapidly as the entire top surface anodizes to form \alo{}. There is a resurgence in the anodization rate as pores begin to form and the exposed surface area increases. Once all pores are formed, the process reaches a roughly steady state and the anodization front progresses through the thickness of the film. The anodization rate drops sharply once the anodization front reaches the bottom of the Al layer, although anodization is continued until the rate begins to level off again, to allow the pores to reach close to the bottom of the Al layer, separated only by a residual barrier layer which is not removed even under further anodization. By varying the anodization voltage, variable pore spacings can be achieved. In this work, voltages of 40 and 57 V were used, resulting in average pore spacings of approximately 100 and 140 nm. After anodization, the pores are just 20~nm in diameter, too small for subsequent pattern transfer by Ar\textsuperscript{+} ion milling~\cite{Mingzhe_AAO}. To widen the pores, we isotropically wet etch using phosphoric acid (5\% aqueous solution) for 56, 75 and 83 minutes at room temperature, obtaining pores of diameter 60, 75 and 80 $\pm $10~nm. A scanning electron microscope (SEM) image of the cleaved edge of the structure (Fig.~\ref{fig_mill}(a)) shows that after anodization and pore widening, the pores terminate about 60~nm above the Nb layer.

To extend the pores through the Nb, we require an anisotropic etch (an isotropic wet etch would completely remove the Nb). We initially tried a CF$_4$ plasma generated by an RF field applied perpendicular to the sample surface (using a parallel plate capacitor geometry to apply the field). This etch would be anisotropic and also chemically selective, etching Nb while not significantly etching \alo{}. SEM images indicated, though, that the Nb was not etched by this process. This may be due to the residual \alo{} at the bottom of the pores (in spite of the phosphoric acid etch) preventing the plasma from reaching the Nb.

We are, however, able to extend the pores using argon ion milling, the effectiveness of which has been demonstrated by milling pores into silicon substrates through AAO~\cite{Mingzhe_AAO}. The cross-section of our film after milling is shown in Fig.~\ref{fig_mill}(b). This technique etches the AAO, and in fact etches the surface more quickly than the bottom of the pores, presumably due to increased likelihood of redeposition of material inside the pores. Thus it is necessary to etch the template until the pores become sufficiently shallow that material is efficiently removed from them. SEM images indicate that the pores do not extend fully through the Nb until the template is thinned to less than 200~nm. Using a beam voltage of 1000~V with an ion current density of 3.9 mA/cm$^2$ (20~mA nominal beam current), we mill for approximately 16 minutes to remove 820~nm of template, leaving 170~nm of template intact. The pores clearly extend through the Nb layer, and when the Nb is peeled away from the SiO$_2$ layer of the substrate, as in Fig.~\ref{fig_mill}(c), the pores are seen to penetrate through into this substrate layer. We speculate that ion milling followed by CF$_4$ etching could be used to produce through-thickness pores in much thicker films.

\begin{figure}
  \includegraphics[width=3.4in]{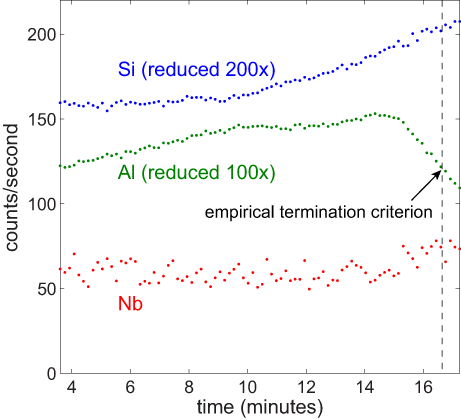}
  \caption{\label{fig_sims}(Color online) Secondary ion mass spectrometry provides a reliable trigger for terminating the milling process. Once the \alo{} layer is sufficiently thin, the Al rate decreases rapidly, accompanied by an increase in the Nb rate as expected.}
\end{figure}

Milling for a fixed time with nominally identical beam parameters was not sufficiently reproducible to reliably remove the desired amount of template without damaging the entire Nb layer. We overcame this problem using secondary ion mass spectrometry to monitor the sputter products during ion milling, as shown in Fig.~\ref{fig_sims}. We were able to detect a downward kink in the milling rate of the Al from the template, accompanied by a slight increase in the Nb milling rate. By examining the cross section of milled samples with SEM, we established an empirical criterion of a 25\% reduction in Al milling rate for terminating the milling with approximately 170~nm of template remaining. The increase in Nb rate during the last 1.7~minutes (90~nm of template milling) corroborates the idea that the template must be thinned to less than 200~nm to effectively mill the Nb. In retrospect, then, the 990~nm template is unnecessarily thick.

\section{\label{Jc}Highly improved \jc{} with tunable matching field}

\begin{figure}
  \includegraphics[width=3.4in]{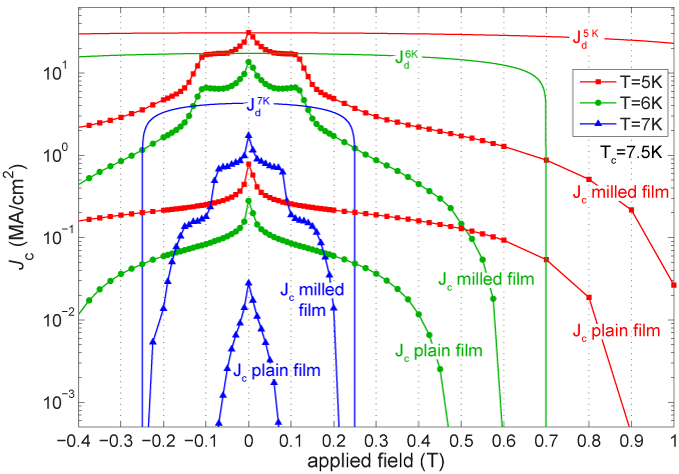}
  \caption{\label{fig_jd}(Color online) Milling pores (75~nm diameter, 140~nm spacing) in a Nb film dramatically improves its transport critical current, compared with an unanodized control. Measurements were taken at 5~K, 6~K and 7~K (\tc{} = 7.5~K). Within each set of curves, the thin solid line is the calculated depairing critical current, $J_{d}$, the thick solid line with points is the porous film, and the dashed line with points is the control.}
\end{figure}

Milling the Nb layer through the AAO template results in a dramatic increase in critical current. The results shown in Fig.~\ref{fig_jd} are obtained from transport measurements on a 12~$\mu$m wide by 10~$\mu$m long bridge patterned by focused ion beam. The magnetic field $H$ is applied perpendicular to the film surface. At 5~K, the milled sample's \jc{} is more than 50 times higher than that of a non-anodized control sample. This calculation of \jc{} does not even take into account the 30\% reduction in film cross section resulting from the introduction of the pores. Thus the artificial pins created by the pores dominate over the natural pinning sites present in an as-deposited film.

This \jc{} enhancement cannot be dismissed as merely poor performance of the control sample. Multiple control samples from the same deposition run were compared and their \jc{} values found to differ by at most 30\%, typically 5\%. Between deposition runs, \jc{} values differed by up to a factor of three, but these differences can be accounted for by corresponding variations in \tc{}, which is known to be extremely sensitive to the background pressure of oxygen during deposition.

What is more remarkable is that at 5~K (\tc{}=7.5~K), the critical current at zero field reaches the calculated Ginzburg-Landau depairing current \cite{Tinkham},
\begin{equation}
 \label{jd}
 \jd{}(T,H) = \frac{2}{3\mu_0\lambda(T)^2}\sqrt{\frac{(\hctwo{}(T)^2-H^2)^{\frac{1}{2}}\Phi_0}{6\pi}}
\end{equation}
where $\mu_0 = 4\pi\times10^{-7}$~H/m is the permeability of free space, $\Phi_0=h/2e=2.07\times10^{-15}$~Wb is the flux quantum, and the upper critical field \hctwo{} is taken from the data as the field at which the IV response becomes linear (Ohmic), corresponding to the transition into the normal state. The penetration depth at 0~K for a film of this thickness in accordance with dirty limit BCS theory, $\lambda(0) = 120$~nm is taken from Ref.~\cite{lambda_value} and scaled for temperature according to the two-fluid model \cite{Tinkham},
\begin{equation}
 \lambda(T) = \frac{\lambda(0)}{\sqrt{1-\left(T/\tc{}\right)^4}}
\end{equation}
This astounding pinning performance is only expected when each vortex sits on an ideal, columnar hole. Although we did not obtain this performance consistently across samples (the best sample is shown), it is impressive that the sample approaches this limit at all. Such high performance only appears in one recent work, also on nanostructured films \cite{nanomesh}.

\begin{figure}
  \includegraphics[width=3.4in]{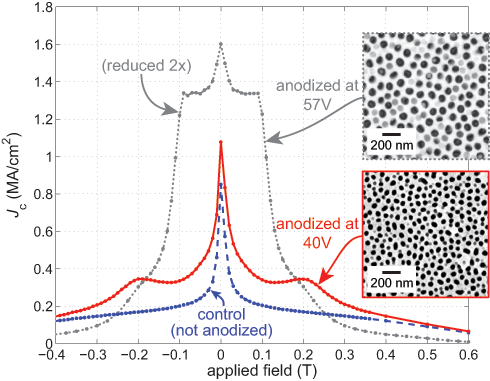}
  \caption{\label{fig_matching_field}(Color online) Films anodized at different voltages have different pore densities (insets) which in turn result in different matching fields in critical current versus magnetic field. (57~V: 75~nm diameter, 140~nm spacing; 40~V: 50 nm~diameter, 100~nm spacing.) The measurement temperature was 6~K; \tc{} = 7.2~K, 8.0~K, and 8.0~K for the 57~V, 40~V, and control films, respectively. \jc{} values of the 57~V sample have been plotted reduced by a factor of two for ease of comparison with the other data.}
\end{figure}

\jc{} shows significant additional enhancement around $0.10\pm0.01$~T and multiples thereof (most evident at 7~K or on the linear scale of Fig.~\ref{fig_Co_Jc}). This matching field represents a filling of one vortex per pore, which accords with the pore density calculated from SEM images of the film surface, counting pores over a known area.
\begin{equation}
 58\pm4 \textrm{ $\mu$m}^{-2} \cdot \Phi_0 = 0.12\pm0.01\textrm{ T}
\end{equation}
The matching field is unchanged between 5 and 6~K, but shifts significantly to lower fields at the highest measurement temperature, 7~K. At this temperature, the coherence length, $\xi$ (calculated as $\xi = \sqrt{\Phi_0/\hctwo{}}$) increases to 90~nm (from 40 and 50~nm at 5 and 6~K, respectively), becoming comparable to the pore spacing of about 145~nm. The more closely spaced pairs of pores may then begin to act as single pinning sites, decreasing the effective density of pinning sites and thereby lowering the effective matching field.

The matching field can be manipulated by changing the pore density, which is easily controlled via the linear relationship between pore spacing and anodization voltage \cite{Adam_AAO}. Fig.~\ref{fig_matching_field} compares results obtained from transport measurements on bridges 120~$\mu$m wide by 500~$\mu$m long photolithographically patterned into films anodized at 40 and 57~V. In both cases, the pore density calculated from the SEM images is consistent with the observed matching field.

\section{\label{magnetic_fab}Fabrication of Magnetic Pins}

We now use this system of strong artificial pinning centers to test whether magnetic inclusions can improve on the existing core pinning, as discussed in the Introduction. Backfilling the pores with magnetic material provides a comparison between magnetic and non-magnetic pins, without changing the superconducting material or geometry of the pins. We chose to backfill with cobalt, a magnetic material that is straightforward to deposit by several methods, and has a large saturation magnetization, producing a flux density of 1.8~T.

\begin{figure}
  \includegraphics[width=2.8in]{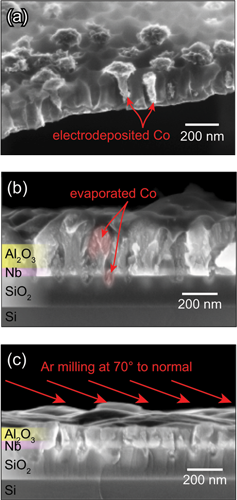}
  \caption{\label{fig_Co}(Color online) To test the effect of magnetization on pinning, we backfill pores with Co. (a) Electrodeposition of Co is non-uniform, filling some pores and not others, as seen in this SEM image. (b) Evaporation results in uniform filling, but also a layer of Co on top of the \alo{}. (c) The top layer of Co is removed by milling at a shallow angle, leaving Co in the pores.}
\end{figure}

Initially, we electrodeposited Co from a cobalt sulfate solution (CoSO$_4\cdot$5H$_2$O 1.3~M, H$_3$BO$_3$ 0.7~M, pH 2.0) at room temperature~\cite{Co_electrodeposition}. Electrodeposition appears to be the ideal method of backfilling the pores, in that the Nb acts as an electrode while the insulating AAO template remains uncoated, so that the film is deposited only at the pinning sites as desired, rather than as a continuous sheet. As illustrated in Fig.~\ref{fig_Co}(a), however, the deposition proved to be very non-uniform, growing in only a third of the pores. Furthermore, many of the Co growths overgrew the template after only 5~seconds of deposition. The deposition was terminated based on a thickness criterion of 150~nm, which translates to a charge (charge is the product of volume, atomic density of Co, and charge per atom); the charge is monitored as the integral of the current during deposition. This termination criterion fixes the deposition volume, so the thickness will increase as the deposition area decreases. The overgrowth is therefore a result of the smaller-than-expected deposition area, and could be corrected by terminating the deposition sooner. However, we found no way of overcoming the non-uniform growth.

We therefore chose, ultimately, to backfill the pores by electron beam evaporation of Co. Evaporation of a 130~nm thick Co layer was performed in high vacuum, leading to directional deposition with low sidewall coverage. This is desirable in order to avoid clogging the pores before they are completely filled. Cross sectional SEM images, as in Fig.~\ref{fig_Co}(b), confirm that evaporation is effective, showing Co (which displays a grain structure that distinguishes it from the AAO and SiO$_2$) present in the bottoms of the pores, penetrating through the Nb layer. As expected, Co is also deposited on top of the template. However, we are able to remove this top layer without affecting the inclusions by off-axis Ar\textsuperscript{+} ion milling. Milling at 70\dg{} from the normal to the substrate with a beam voltage of 1000~V and an ion current density of 3.9 mA/cm$^2$ for five minutes removes the top Co layer and half of the remaining 170~nm of template. A cross section of the final film is shown in Fig.~\ref{fig_Co}(c).

\section{\label{magnetic_jc}Effect of magnetization on critical current}

\begin{figure}
  \includegraphics[width=3.4in]{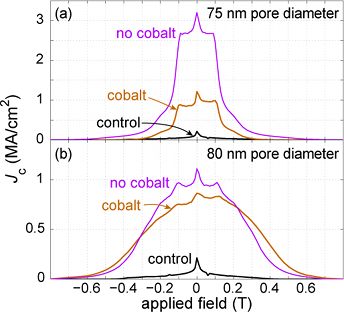}
  \caption{\label{fig_Co_Jc}(Color online) A film backfilled with Co shows lower \jc{} than its unfilled counterpart, possibly due to an adverse proximity effect. Two films are shown with different pore diameters (75 and 80~nm) but the same pore spacing (140~nm), measured at 6~K. An unanodized control sample is also shown. The \tc{} of the control sample is 8.2~K; all others are 7.2~K.}
\end{figure}

Contrary to our initial expectations, adding Co to the pores does not significantly increase \jc{}. To obtain the data in Fig.~\ref{fig_Co_Jc}, films were coated with Co over part of the substrate using a shadow mask during evaporation. Transport bridges 120~$\mu$m wide by 500~$\mu$m long were then photolithographically patterned in each section of the sample, allowing a comparison of magnetic and non-magnetic \jc{} on the same sample. For a film containing 75~nm diameter pores (Fig.~\ref{fig_Co_Jc}(a)), \jc{} decreases across all applied fields with the addition of Co. \tc{} is 7.2~K for both of the bridges, and therefore does not offer an explanation for the decrease. Increasing the pore diameter is expected to decrease the overall \jc{} as the superconducting cross-section is reduced. In itself, this is insufficient to explain the factor of three reduction in \jc{} values on increasing the pore size of the non-magnetic samples. It could, however, explain the slight reduction in self-field \jc{} when comparing the two magnetic samples. The magnetic sample with 80 nm diameter pores shows more closely comparable values to the non-magnetic sample, with even some enhancement at $\left|\mu_0H\right|>0.3$~T, but still a lower value at zero field, where Lorentz force reduction predicts the greatest increase~\cite{Blamire_magnetic_pinning_theory}. The optimal pin size for a vortex core pin is well known to be $\sim$2$\xi$ = 76~nm for Nb -- the size at which the non-magnetic sample exhibits the greatest enhancement. The optimal pin size for magnetic pinning depends on the mechanism at work. For Lorentz force reduction, the optimal pin size would be one that could carry a full quantum of flux. For fully saturated Co ($\mu_0M_s = 1.8$~T), this would be a 38~nm diameter. Based on the actual Co magnetisation loop of these samples, however, an applied field of just above 0.1~T ($\mu_0M \approx 0.5$~T) would be optimal for these pore diameters. At this field, both non-magnetic samples are outperforming their magnetic counterparts, contradicting the predictions of the Lorentz force reduction theory of magnetic pinning. 

There are likely factors other than the magnetism of the Co behind the seemingly disappointing performance of the Co-filled samples compared to their unfilled counterparts. For example, the core pinning potential is significantly reduced by the proximity effect when a conductor, whether magnetic or not, is brought into contact with the superconductor (i.e. the Co deposited in the pores). The gradient in potential energy (and hence core pinning force) would then be determined by the length scale of this proximity effect's suppression of superconductivity, rather than the coherence length. We suggest, then, that depositing a ferromagnetic insulator would allow a more perfect experimental comparison than our ferromagnetic metal.

\begin{figure}
  \includegraphics[width=3.2in]{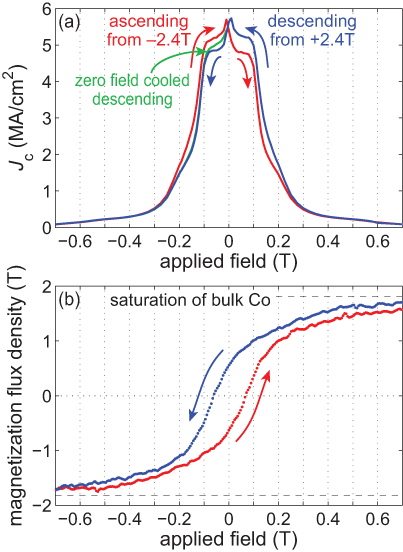}
  \caption{\label{fig_hysteresis}(Color online) (a) Hysteresis in \jc{} versus applied field at 5~K indicates that aligning the magnetization of the Co inclusions parallel to the superconducting vortices increases \jc{}. The hysteresis follows the hysteretic magnetization of the Co, shown in (b). A sample of Co inclusions with the top layer of Co removed is measured at 300~K with the applied field perpendicular to the film surface. Red and blue curves indicate increasing and decreasing applied field, respectively.}
\end{figure}

Despite the overall decrease in \jc{}, there is evidence for a magnetic contribution to pinning in the form of a hysteresis in \jc{} versus applied field that is not present in the non-magnetic samples (Fig.~\ref{fig_hysteresis}(a)). This hysteresis arises from the hysteretic magnetization curve of the Co inclusions. Fig.~\ref{fig_hysteresis}(b) shows the magnetization (at 300~K) of an identical sample that is not patterned into transport bridges. We expect that this magnetization is not significantly different in the presence of superconductivity, since the large demagnetization factor of a thin film in perpendicular field $H$ makes the field inside and outside the sample nearly the same. Furthermore, the internal field is approximately homogeneous when the vortices are spaced more closely than the thin film effective penetration depth (true for $\left|\mu_0H\right| > 2$~mT).

The hysteresis in \jc{} tells us that pinning is stronger when the inclusion magnetization is aligned with the vortices: for positive applied fields, the critical current is higher when descending in applied field (the blue curve in Fig.~\ref{fig_hysteresis}(a)), which corresponds to more positive magnetization (the blue curve in Fig.~\ref{fig_hysteresis}(b)). The opposite is true for negative applied fields. This is qualitatively consistent with the Lorentz force reduction theory~\cite{Blamire_magnetic_pinning_theory}, in which the force acting on a vortex pinned by a magnetic inclusion is diminished by simultaneously acting on the inclusion, but is also consistent with several other mechanisms of magnetic pinning, including field compensation~\cite{RDS_math} whereby the applied magnetic field is to some extent compensated by the field generated by the magnetic material, and magnetic attraction~\cite{cardoso} in which the pinning force acting on the vortex is supplemented by an additional magnetic component.

\begin{figure}
  \includegraphics[width=3.4in]{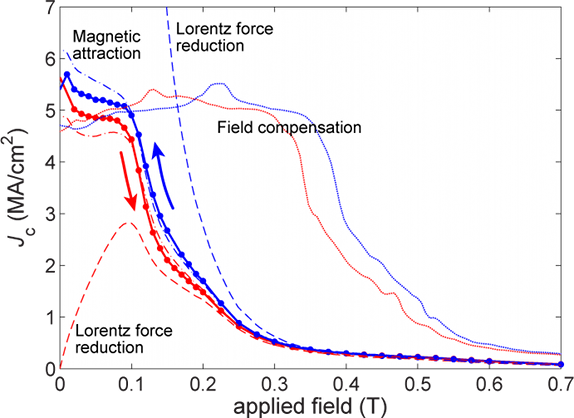}
  \caption{\label{fig_fit}(Color online) The size of the hysteresis in \jc{} (data shown as a solid line with points; red curves correspond to increasing applied field, and blue to decreasing applied field) allows comparison with the calculated predictions of several theories of magnetic pinning. The large difference between \jc{} on ascending and descending sweeps of the applied field expected from Lorentz force attraction (dashed curves) is not observed, nor is the substantial field shift predicted by the field compensation theory (dotted curves). The data is most consistent with the simple magnetic attraction calculation (dash-dot curves).}
\end{figure}

To distinguish between these theoretical mechanisms, we must look quantitatively at the size of the hysteresis. In Fig.~\ref{fig_fit}, we have re-plotted the hysteretic \jc{} versus $H$ for $H>0$. We first attempt to fit the Lorentz force reduction theory to these data using Eqs.~8 and 10 of Ref.~\cite{Blamire_magnetic_pinning_theory}:

\begin{equation}
 \frac{\jc{}'}{\jc{}} = (1-f) \frac{B}{B'}
 \label{eq:Lorentz_force_reduction}
\end{equation}
\begin{equation}
 B = B' + \mu_0 M(B') \cdot f
\end{equation}
where $\jc{}'/\jc{}$ is the ratio between the magnetic and non-magnetic critical current densities, $f$ is the volume fraction of the inclusions (equal to the area fraction of the pores, measured from SEM images to be 0.16)\footnote{Since the pores are uniform through the thickness of the film, the area fraction of pores is given by the fraction of a line drawn within the plane that intersects the pores. Unless the distribution of pores is correlated with the sampling subvolume (the plane in the case of the area fraction), this is equivalent to a random sampling of the entire volume, yielding the volume fraction.}, $B$ is the applied magnetic field, and $B'$ is an effective field. This effective field is calculated by numerically inverting the second equation, using the measured magnetization curve of the inclusions, $M(H)$. Having calculated the ratio $\jc{}'/\jc{}$ as a function of $H$, we multiply it by the average of the measured ascending and descending \jc{} curves as an approximation of the form of the non-magnetic \jc{}. (As explained above, the non-cobalt filled sample proved unsuitable for direct comparison with this sample.) The prediction of the theory in this case is not the absolute \jc{}, since we have no direct measurement of the non-magnetic \jc{}, but it does yield the ratio between ascending and descending curves. The measured data cannot satisfy the large predicted divergence of the ascending and descending curves, however. In short, Lorentz force reduction theory predicts a much larger effect of magnetization on \jc{} than that observed.

The equivalent prediction for the field compensation scenario~\cite{RDS_math} requires transforming the field axis, $H \rightarrow H'$, according to the measured magnetization of the inclusions.
\begin{equation}
H' = H - fM(H)
\label{eq:field_compensation}
\end{equation}
where, as before, $f$ is the area fraction of the inclusions. The second term represents the spatially averaged return field of the inclusions, which is subtracted from the applied field. As above, we employ the average of the ascending and descending data as the starting point for the transformation. The result is plotted as the solid lines in Fig.~\ref{fig_fit}. The hysteresis in the calculated curves follows from the hysteresis in the measured $M(H)$. The large shift in field predicted by this model is also not consistent with the observed \jc{} behavior.

Finally, we consider the magnetic attraction between vortices and magnetized inclusions~\cite{cardoso}. The pinning potential per length of vortex is modified by the addition of the Zeeman energy term $-AM\tilde{H}/2$, where $A$ is the area of the inclusion perpendicular to the applied field direction, $M$ is its magnetization, and $\tilde{H}$ is the field applied to the inclusion by the vortex lattice. The additional magnetic pinning force is then the gradient of this additional pinning potential. Both $M$ and $\tilde{H}$ vary as a function of $H$, the macroscopic applied field. We took $M(H)$ to be as shown in Fig.~\ref{fig_hysteresis}(b). To estimate $\tilde{H}$, we summed the field from two neighboring vortices at a point between them a distance $r$ from one of them:
\begin{equation}
\tilde{H}(r)=\frac{\Phi_0}{2\pi\mu_0\lambda^2} \left[ K_0\left(\frac{r}{\lambda}\right) + K_0\left(\frac{a-r}{\lambda}\right) \right]
\end{equation}
where $K_0$ is a modified Bessel function, and $a$ is the expected vortex lattice spacing for the particular applied field,
\begin{equation}
a=\sqrt{\frac{2\Phi_0}{\mu_0H\sqrt{3}}}
\end{equation}

We then calculated the magnetic pinning force, $f_p^{mag}$, as the gradient of the resulting potential, evaluating the result with the inclusion offset from the center of the vortex by a distance $\xi$, as this is the position we expect a pinned vortex to occupy due to the strong core pinning of the inclusions:
\begin{equation}
f_p^{mag}=\frac{AM}{2}\left[\frac{d\tilde{H}}{dr}\right]_{r=\xi}
\end{equation}
Below the first matching field (i.e. at vortex densities lower than one vortex per pore), we assume each vortex is paired with one inclusion, thus the additional pinning force above contributes directly to a change in critical current, $\Delta\jc{} = f_p^{mag}/\Phi_0$. Above this field, we applied the additional magnetic pinning force to the vortex lattice as a whole, dividing the single vortex pinning force by the ratio of the density of vortices to the density of pores.

To obtain the magnetic attraction (dash-dot) curves shown in Fig.~\ref{fig_fit}, we add the calculated magnetic pinning force (which is positive when $M$ and $H$ are aligned, and negative when they are anti-aligned) to the average of the ascending and descending sweeps from the measured data. Thus we are comparing the difference between ascending and descending \jc{} curves in the data and the calculation. They are in fact comparable, particularly above the first matching field. Below the first matching field, the calculation yields a larger magnetic contribution to pinning than that observed. Idealizations in the calculation such as the placement of the vortices entirely on inclusions may account for the disparity.

\section{\label{conclusion}Conclusions}

We have exploited the scalable, self-assembled nanofabrication technique of porous anodized alumina to make through-thickness pores in a superconducting film. These pores act as highly effective artificial pinning centers with a tunable size and density. With this system, we are able to unambiguously demonstrate and quantify the effect of magnetization on pinning, observing a unique hysteretic \jc{}$(H)$ characteristic stemming from the hysteresis of the inclusion magnetization. We find, however, that the size of this effect is smaller than predicted by the Lorentz force reduction theory. This is not unexpected, primarily because the theory is computed for an ideally soft ferromagnet whose magnetization is proportional to applied field whereas the Co nanopillars have sufficient coercivity to prevent their magnetization from being realigned by individual vortices. A field compensation theory similarly fails to describe the observed behavior. Instead, our magnetic pinning is best modeled by considering the magnetic attraction between the vortex lattice and the inclusion lattice.

\begin{acknowledgments}
We thank Dr Katherine Develos-Bagarinao, Dr Suman-Lata Sahonta and Dr Nadia Stelmashenko for assistance with this work. RBD thanks Hughes Hall, Cambridge for a research fellowship. SCW is supported by The Leverhulme Trust, with supplementary funding from The Isaac Newton Trust. This work was supported by the UK Engineering and Physical Sciences Research Council (grant number E011020).
\end{acknowledgments}

\clearpage

\providecommand{\newblock}{}

\end{document}